\documentclass[aps,prb,twocolumn,superscriptaddress,showpacs]{revtex4-1}
\usepackage{dcolumn} 
\usepackage{graphicx}
\usepackage{bm}      
\usepackage{color}
\usepackage{amsmath,amssymb}
\usepackage{hyperref}
\usepackage{multirow}
\usepackage{epstopdf}
\usepackage{latexsym}
\usepackage{subfigure}
\usepackage{wasysym} 
\usepackage{times}
\usepackage{gensymb}
\usepackage{textcomp}

\begin{document}

\title{Ba$_8$MnNb$_6$O$_{24}$: a model two-dimensional spin-5/2 triangular lattice antiferromagnet}

\author{R.~Rawl}
\affiliation{Department of Physics and Astronomy, University of Tennessee, Knoxville, Tennessee 37996, USA}

\author{L.~Ge}
\affiliation{School of Physics, Georgia Institute of Technology, Atlanta, GA 30332, USA}

\author{Z.~Lu}
\affiliation{Helmholtz-Zentrum Berlin f\"ur Materialien und Energie, D-14109 Berlin, Germany}

\author{Z.~Evenson}
\affiliation{Heinz Maier-Leibnitz Zentrum (MLZ) and Physik Department, Technische Universit\"at M\"nchen, Garching 85748, Germany}

\author{C.~R.~Dela Cruz}
\affiliation{Quantum Condensed Matter Division, Oak Ridge National Laboratory, Oak Ridge, Tennessee 37381, USA}

\author{Q.~Huang}
\affiliation{Department of Physics and Astronomy, University of Tennessee, Knoxville, Tennessee 37996, USA}

\author{M.~Lee}
\affiliation{Department of Physics, Florida State University, Tallahassee, FL 32306, USA}
\affiliation{National High Magnetic Field Laboratory, Florida State University, Tallahassee, FL 32310, USA}

\author{E.~S.~Choi}
\affiliation{National High Magnetic Field Laboratory, Florida State University, Tallahassee, FL 32310, USA}

\author{M. Mourigal}
\affiliation{School of Physics, Georgia Institute of Technology, Atlanta, GA 30332, USA}

\author{H.~D.~Zhou}
\email{hzhou10@utk.edu}
\affiliation{Department of Physics and Astronomy, University of Tennessee, Knoxville, Tennessee 37996, USA}
\affiliation{Key Laboratory of Artificial Structures and Quantum Control, School of Physics and Astronomy, Shanghai Jiao Tong University, Shanghai 200240, China}

\author{J. Ma}
\email{jma3@sjtu.edu.cn}
\affiliation{Key Laboratory of Artificial Structures and Quantum Control, School of Physics and Astronomy, Shanghai Jiao Tong University, Shanghai 200240, China}
\affiliation{Collaborative Innovation Center of Advanced Microstructures, Nanjing University, Nanjing, Jiangsu 210093, China}

\date{\today}

\begin{abstract}
We successfully synthesized and characterized the triangular lattice anitferromagnet Ba$_8$MnNb$_6$O$_{24}$, which comprises equilateral spin-5/2 Mn$^{2+}$ triangular layers separated by six non-magnetic Nb$^{5+}$ layers. The detailed susceptibility, specific heat, elastic and inelastic neutron scattering measurements, and spin wave theory simulation on this system reveal that it has a 120 degree ordering ground state below $T_{\text{N}}$ = 1.45 K with in-plane nearest-neighbor exchange interaction $\approx$ 0.11 meV. While the large separation 18.9\AA\ between magnetic layers makes the inter-layer exchange interaction virtually zero, our results suggest that a weak easy-plane anisotropy is the driving force for the $\mathbf{k}_\mathrm{m} = (1/3,1/3,0)$ magnetic ordering. The  magnetic properties of Ba$_8$MnNb$_6$O$_{24}$, along with its classical excitation spectra, contrast with the related triple perovskite Ba$_3$MnNb$_2$O$_{9}$, which shows easy-axis anisotropy, and the iso-structural compound Ba$_8$CoNb$_6$O$_{24}$, in which the effective spin-1/2 Co$^{2+}$ spins do not order down to 60 mK and in which the spin dynamics shows sign of strong quantum effects.
\end{abstract}
\maketitle

\section{INTRODUCTION}

Truly two dimensional (2D) lattices of interacting spins, including triangular, honeycomb, and kagome antiferromagnets, are of central interest to stabilize, explore and understand exotic quantum states and their excitations \cite{YMG1,YMG2,YMG3,RuCl1,RuCl2,RuCl3,ZnCu1,ZnCu2,ZnCu3,TKL1,TKL2,CsCuBr1,CsCuBr2,CoSb1,CoSb2,CoSb3,CoSb4,CoSb5,CoSb6,NiSb,CoNb1,CoNb2}. However, the  experimental realization of an ideal 2D magnetic system embedded in a bulk crystal is very difficult, since undesired factors such as lattice distortions, inter-plane interactions, and anisotropies are often present and transform the system of interest into, at best, a quasi-2D environment.

\begin{figure}
	\linespread{1}
	\par
	\begin{center}
		\includegraphics[width= 3.5 in]{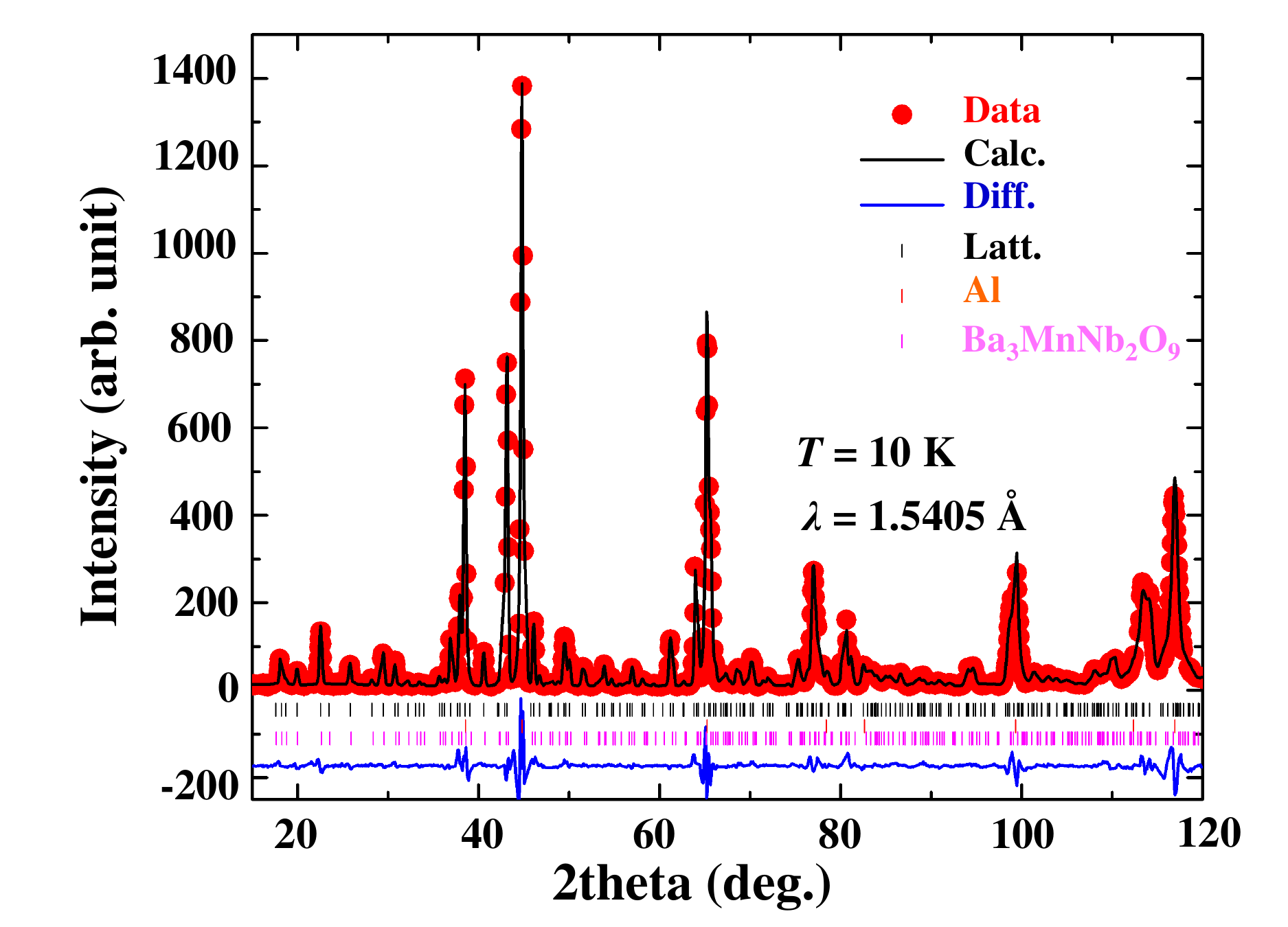}
	\end{center}
	\par
	\caption{\label{Fig:1} (color online)  Rietveld refinement of the neutron powder diffraction pattern of Ba$_8$MnNb$_6$O$_{24}$ measured at $T$ = 10 K with $\lambda$ = 1.5405 \AA. Red circles are the measured intensity, the black line is the calculated intensity, and the blue line is the difference. Tick marks indicate lattice Bragg peak positions for Ba$_8$MnNb$_6$O$_{24}$ (upper black marks), the aluminum cab background (middle red marks), and $\sim$ 3$\%$ Ba$_3$MnNb$_2$O$_{9}$ impurity (lower magenta marks).
}
\end{figure}
\begin{figure}
	\linespread{1}
	\par
	\begin{center}
		\includegraphics[width= 3.4 in]{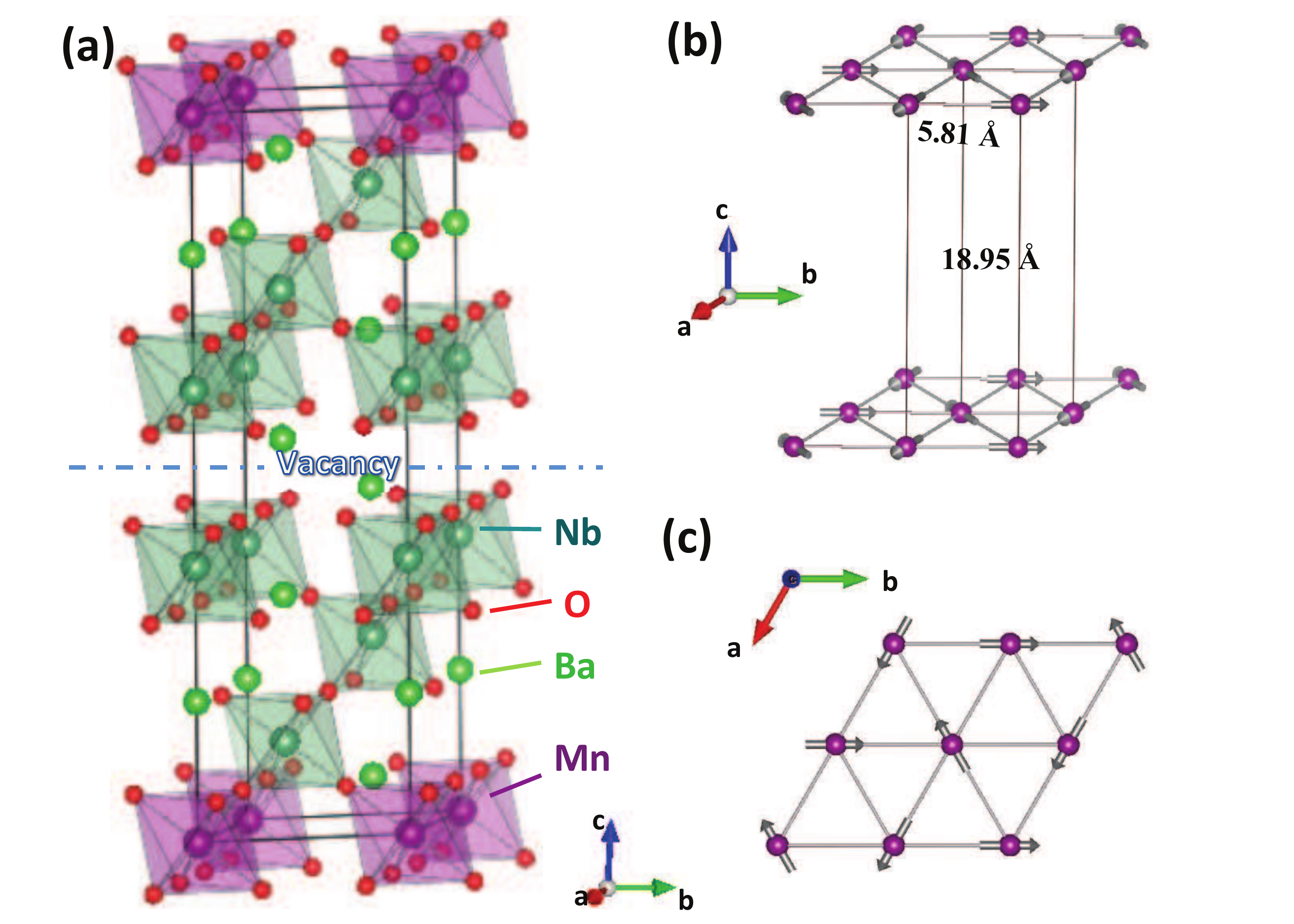}
	\end{center}
	\par
	\caption{\label{Fig:2} (color online) (a) Stacked layer structure of Ba$_8$MnNb$_6$O$_{24}$. (b) Unit cell of Ba$_8$MnNb$_6$O$_{24}$ and spin structure of Mn$^{2+}$ ions at zero field.  (c) Triangular lattice composed of Mn$^{2+}$ ions in $ab$ plane.
}
\end{figure}
The design and synthesis of bulk materials with an ideal 2D magnetic lattice has been a challenge in the materials science community. Here, our strategy to achieve two dimensionality is to insert nonmagnetic buffer layers between magnetic layers in a three-dimensional crystal structure. This idea has been applied to the perovskite structure (ABO$_3$) comprised of stacked triangular layers of B ions to yield triple perovskite structure (A$_3$B'B$_2$O$_9$). By having triangular layers of magnetic B' ions separated by two triangular layers of nonmagnetic B ions, the inter-layer distance gets increased and the inter-layer interactions weakened. As a result, A$_3$B'B$_2$O$_9$ realizes ideal quasi-2D magnets for exploring exotic magnetic properties. Examples of such a triple perovskite structure are Ba$_3$B'Nb$_2$O$_9$ and Ba$_3$B'Sb$_2$O$_9$ (B'$^{2+}$ = Co$^{2+}$, Ni$^{2+}$, and Mn $^{2+}$ with spin number 1/2, 1, and 5/2, respectively). \cite{MnSb1,MnNb,NiNb,NiSb,CoNb1,CoNb2,CoSb1,CoSb2}

Among these materials, Ba$_3$CoSb$_2$O$_9$ is the arguably most interesting one stabilizing a one-third magnetization plateau as well as carrying anomalous zero-field magnetic excitations. \cite{CoSb1,CoSb2,CoSb3,CoSb4,CoSb5,CoSb6,KamiyaNC2018, ItoNC2017, KamiyaNC2018} Proceeding from the triple perovskite structure, and with the strategy outlined above, we previously modified and expanded Ba$_3$CoSb$_2$O$_9$ to Ba$_8$CoNb$_6$O$_{24}$ with six layers of nonmagnetic Nb and a vacant layer between each layer of magnetic Co$^{2+}$ ions. \cite{Ba8Co1,YuPRM2018} In Ba$_3$CoSb$_2$O$_9$, the intra-plane Co--Co distance is  5.86 \AA \, and the inter-plane Co--Co distance is 7.23 \AA\, which yields an inter-layer exchange interaction ($J'$) around 5\% the strength of the intra-layer exchange  interaction($J$). \cite{CoSb1,CoSb2,CoSb3,CoSb4,CoSb5,CoSb6,KamiyaNC2018, ItoNC2017, KamiyaNC2018} Also present in Ba$_3$CoSb$_2$O$_9$ is a small easy-plane XXZ anisotropy (the ratio between the longitudinal and transverse exchange interactions is $\Delta\!\approx\!0.9$). In contrast,  Ba$_8$CoNb$_6$O$_{24}$ has similar intra-layer Co--Co distance of 5.79 \AA \, while the inter-layer distances expand to 18.90 \AA. This reduces the inter-plane interaction below detectable limits of susceptibility and specific heat measurements and also removes any resolvable anisotropy, producing a virtually ideal 2D magnetic lattice with no ordering down to 60 mK, as our recent studies show. Moreover, its inelastic neutron scattering spectrum reveals a high-energy continuum also known as two-magnon scattering, which reflects the reduction of the ordered moment by quantum fluctuations. Therefore, Ba$_8$CoNb$_6$O$_{24}$ is a rare example of spin-1/2 triangular-lattice Heisenberg antiferromagnet in the 2D limit.\cite{Ba8Co1,YuPRM2018}

Following upon the successful reduction of $J'$ and anisotropy in Ba$_8$CoNb$_6$O$_{24}$, calls for engineering a similar material with classical spins. In this paper, we report the synthesis and characterization of Ba$_8$MnNb$_6$O$_{24}$, which is isostructural to Ba$_8$CoNb$_6$O$_{24}$, but with a  Mn$^{2+}$ (spin-5/2) 2D triangular lattice. The detailed DC, AC susceptibility, specific heat, elastic and inelastic neutron scattering measurements on this system reveal that it has a 120 degree ordering ground state below $T_{\rm N}$ = 1.45 K. The linear spin-wave simulation along with the inelastic neutron scattering (INS) spectra extracts the antiferromagnetic nearest neighbor exchange $J = 0.11$ meV. Unlike the related triple perovskite Ba$_3$MnNb$_2$O$_{9}$ that shows the easy-axis anisotropy,\cite{MnNb} we suggest Ba$_8$MnNb$_6$O$_{24}$ could have a rather weak easy-plane anisotropy.

Our paper is organized as follows. Sec.~II presents our experimental and theoretical methods. Sec.~III presents the results of our thermo-magnetic, diffraction, and inelastic neutron scattering characterizations of powder samples of Ba$_8$MnNb$_6$O$_{24}$. Sec.~IV discusses and interprets our experimental results. Sec.~V serves as a conclusion.

\section{EXPERIMENTAL}

Our polycrystalline sample of Ba$_8$MnNb$_6$O$_{24}$ was prepared by solid state reaction. Stoichiometric amounts of BaCO$_3$, MnO, and Nb$_2$O$_5$ were mixed in agate mortars, compressed into pellets, and annealed for 20 hours at temperatures of 1525 $^\circ$C and 1600 $^\circ$C under Ar atomosphere with intermediate mixing. High-resolution neutron powder diffraction (NPD) measurements were performed by a neutron powder diffractometer, HB2A, at the High Flux Isotope Reactor (HFIR), Oak Ridge National Laboratory (ORNL), USA. Around 3 grams of powder was loaded in an Al-cylinder can and mounted in a close-cycled refrigerator. We used a neutron wavelength of $\lambda$ = 1.5405 and 2.4127$\, $\AA $\,$ with a collimation of 12$^\prime$-open-6$^\prime$. The NPD patterns were analyzed by the Rietveld refinement program FullProf \cite{Fprof}. Powder inelastic neutron scattering was measured using the cold neutron time-of-flight spectrometer (TOFTOF) at the Heinz Maier-Leibnitz Zentrum (MLZ), Munich, Germany. Around 6 grams of powder was loaded in an Al cylindrical can mounted at the bottom of a dilution refrigerator. An incident neutron energy of $E_i\!=\!3.27$~ meV was used what yielded an elastic energy resolution of 0.08 meV. The data were binned into steps of 0.015$\, $\AA $\,$ and 0.015 meV. The DC magnetic susceptibility measurements were performed using a Quantum Design superconducting interference device (SQUID) magnetometer with an applied field of 0.5 T.  The DC magnetization was performed on a vibrating sample system (VSM) at the National High Magnetic Field Laboratory (NHMFL), Tallahassee, USA. The AC susceptibility measurements were taken using with the conventional mutual inductance technique with a homemade setup\cite{AC}. The specific heat data were obtained using a commercial physical property measurement system (PPMS, Quantum Design).

\section{RESULTS}

\subsection{Lattice structure}

\begin{table*}[tp]
\par
\caption{Structural parameters for Ba$_8$MnNb$_6$O$_{24}$ at 10 K (space group \textit{P-3m1}) determined from refined NPD measurements. }
\label{Tab:1}
\setlength{\tabcolsep}{0.55cm}
\begin{tabular}{ccccccc}
\hline
\hline
	Refinement & Atom & Site & {\it x} & {\it y} & {\it z} & Occupancy \\ \hline
	\multirow{5}{*}{\begin{tabular}[c]{@{}c@{}}\\ Ba$_8$MnNb$_6$O$_{24}$\\ RF-factor = 6.30 \\ Bragg $R$-factor = 7.28 \\ \end{tabular}} & Ba1 & 2c &$   $ 0 & 0 & 0.18732(84) & 0.16666 \\
	 & Ba2 & 2c &$   $ 1/3 & 2/3 & 0.05867(120) & 0.16666 \\
	 & Ba3 & 2c &$   $ 1/3 & 2/3 & 0.44997(89) & 0.16666 \\
	 & Ba4 & 2c &$   $ 1/3 & 2/3 & 0.68171(92) & 0.16666 \\
	 & Mn & 1a &$   $ 0 & 0 & 0 & 0.08333 \\
	 & Nb1 & 2c &$   $ 0 & 0 & 0.38767(61) & 0.16667 \\
	 & Nb2 & 2d &$   $ 1/3 & 2/3 & 0.25790(82) & 0.16667 \\
	 & Nb3 & 2d &$   $ 1/3 & 2/3 & 0.87545(74) & 0.16667 \\
	 & O1 & 6i &$   $ 0.17214(116) & 0.82786(116) & 0.30799(46) & 0.50 \\
	 & O2 & 6i &$   $ 0.16375(92) & 0.83615(92) & 0.57010(49) & 0.50 \\
	 & O3 & 6i &$   $ 0.16731(117) & 0.83259(117) & 0.93445(58) & 0.50 \\
	 & O4 & 6i &$   $ 0.49493(118) & 0.50497(118) & 0.18831(31) & 0.50 \\
\hline
	\multicolumn{1}{l}{} & \multicolumn{1}{l}{} & \multicolumn{1}{l}{} & \multicolumn{1}{l}{} & \multicolumn{1}{l}{} & \multicolumn{1}{l}{} & \multicolumn{1}{l}{} \\
Space group	& \multicolumn{4}{c}{ Lattice parameters ({\AA})} &  &  \\
$P$ - 3 $m$ 1  & \multicolumn{4}{c}{$a$ = $b$ = 5.80701(6), $c$ = 18.94654(29)} & \multicolumn{2}{l}{Overall B-factor = 0.156 ({\AA}$^{2}$)}\\
Magnetic space group	& \multicolumn{4}{c}{Momentum}  & \\
$P$ - $ 1$  & \multicolumn{4}{c}{4.05(35) $\mu_B$ } & & \\
	\multicolumn{1}{l}{} & \multicolumn{6}{l}{} \\
	\hline
\end{tabular}
\end{table*}
Figure~\ref{Fig:1} shows the NPD pattern of Ba$_8$MnNb$_6$O$_{24}$ measured at $T=10$~K with wavelength $\lambda$ = 1.5405 \AA. A Rietveld refinement yields a pure phase of the space group \textit{P-3m1} with the lattice parameters of $a$ = 5.8070(1) \AA\ and $c$ = 18.9465(3) \AA. The details of the structural parameters are listed in Table~\ref{Tab:1} and confirm that Ba$_8$MnNb$_6$O$_{24}$ is isostructural to Ba$_8$CoNb$_6$O$_{24}$. No anti-site disorder between Mn$^{2+}$ and Nb$^{5+}$ ions was observed, while $\sim$ 3$\%$ Ba$_3$MnNb$_2$O$_{9}$ had been identified as the impurity. A crystalographic unit-cell [Fig.~\ref{Fig:2}(a)] contains a vacant layer and \emph{six layers} of non-magnetic NbO$_6$ octahedral separating triangular layers of Mn$^{2+}$ ions . While the intra-layer Mn--Mn distance is around 5.81~{\AA}, the inter-layer Mn--Mn distance is as large as 18.91~{\AA} [Fig.~\ref{Fig:2}(b)]. This remarkable structure is expected to guarantee that the inter-layer interaction of Ba$_8$MnNb$_6$O$_{24}$ is approaching the zero limit.

\subsection{DC and AC susceptibility}
\begin{figure}
	\linespread{1}
	\par
	\begin{center}
		\includegraphics[width= 3.2 in]{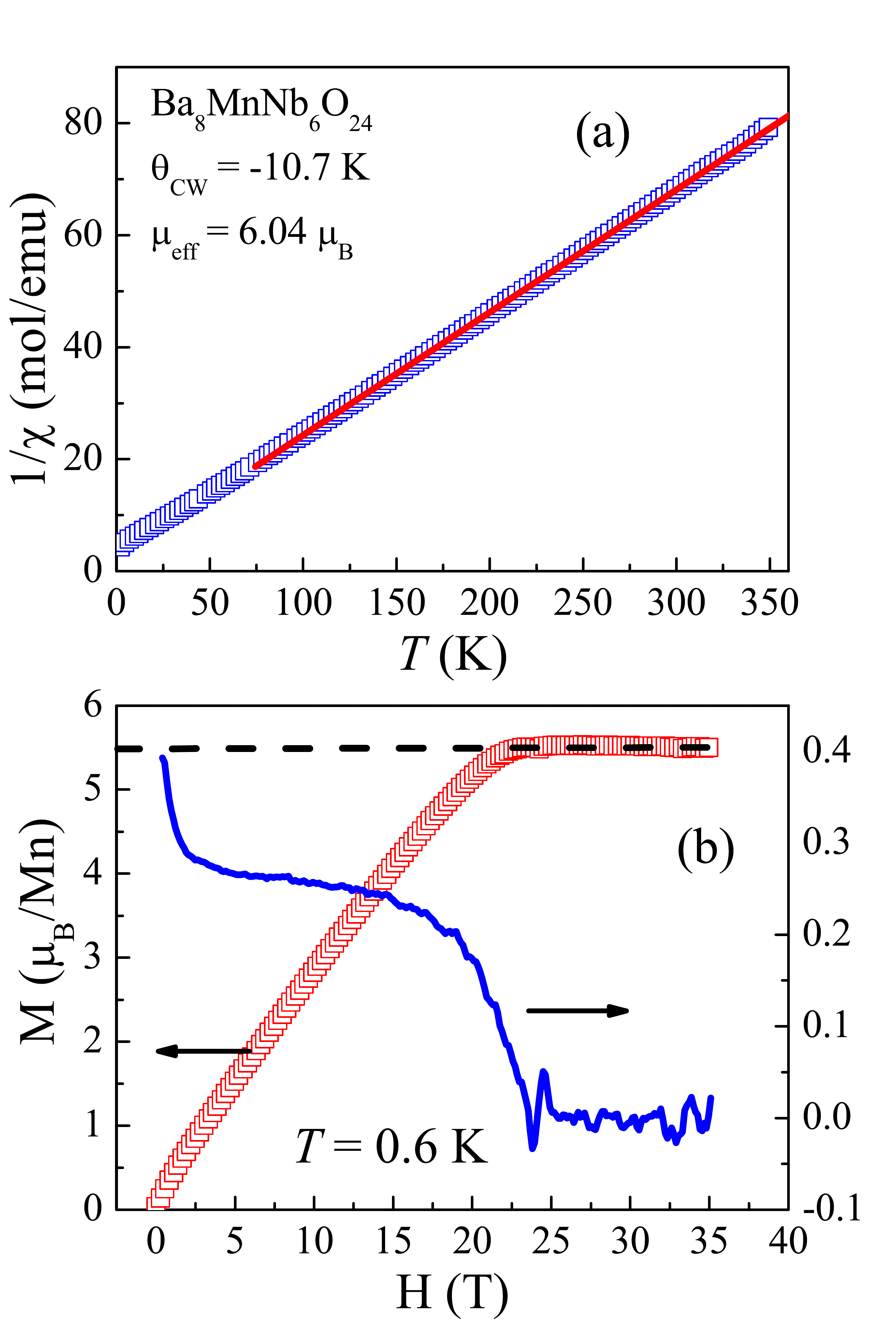}
	\end{center}
	\par
	\caption{\label{Fig:3} (color online) (a) The inverse of the DC susceptibility (open squares) measured at a field of 0.5 T. The solid line is the Curie-Weiss fitting from 100 $-$ 350 K. (b) The DC magnetization up to 35 T measured at 0.6 K,  showing a saturation field $\sim$22.5 T with a saturation moment of 5.50 $\mu_{\rm B}$/Mn$^{2+}$. The solid line is the derivative of the magnetization.
}
\end{figure}

Figure~\ref{Fig:3}(a) reports the temperature dependence of the magnetic DC susceptibility $\chi$ of Ba$_8$MnNb$_6$O$_{24}$. No sign of magnetic ordering is observed down to 1.8 K and a Curie-Weiss fit of the inverse DC susceptibility on the range of temperatures 100-350 K yields $\mu_{\rm eff}\!=\!6.04$~$\mu_{\rm B}$ and $\theta_{\rm W}\!=\!-10.7$~K. The negative Weiss constant indicates overall antiferromagnetic exchange interactions, while the obtained effective magnetic moment agrees well with the value 5.93 $\mu_{\text{B}}$ expected for spin-only $S\!=\!5/2$ magnetic moments \cite{effectivem1}. Hereafter, we will consider the Hamiltonian of a triangular Heisenberg antiferromagnet,
\begin{align}
    \mathcal{H} = J\sum\limits_{\left<i,j\right>}\left(S^x_i S^x_j + S^y_i S^y_j + \Delta S^z_i S^z_j \right),
    \label{eq:ham}
\end{align}
where $\left<i,j\right>$ indicates the nearest neighbors and $\Delta$ indicates the potential easy-plane anisotropy which is not considered until Section III.D. Considering $z=6$ nearest-neighbour magnetic moments coupled with the Heisenberg exchange interaction $J$, mean-field theory yields $\theta_{\text{W}}=-zJS(S+1))/6k_{\rm{B}}$, what corresponds to $J/k_{\rm{B}}\!=\!-4/35~\theta_{\text{CW}}$ = 1.22 K for Ba$_8$MnNb$_6$O$_{24}$. The isothermal DC magnetization taken at $T\!=\!~$0.6 K is shown in \ref{Fig:3}(b) and yields a moment of 5.5 $\mu_{\rm B}$ above the saturation magnetic field $\mu_0 H_\mathrm{s}\!\approx\!22.5$~T.  The value of the saturated moment corresponds to a powder-averaged gyromagnetic ratio $g$ = 2.1 for Mn$^{2+}$ ion assuming $S=5/2$. Therefore, the exchange interaction, $J$ = ${\frac {g \mu_B H_{sat} S^{-1} } {9} } $ =1.41 K, from the saturated field agreed with the temperature-dependence measurement of magnetization. The field-derivative of the magnetization shows no obvious valley or peak that would indicate possible spin-state transitions.

\begin{figure}
	\linespread{1}
	\par
	\begin{center}
		\includegraphics[width= 3.2 in]{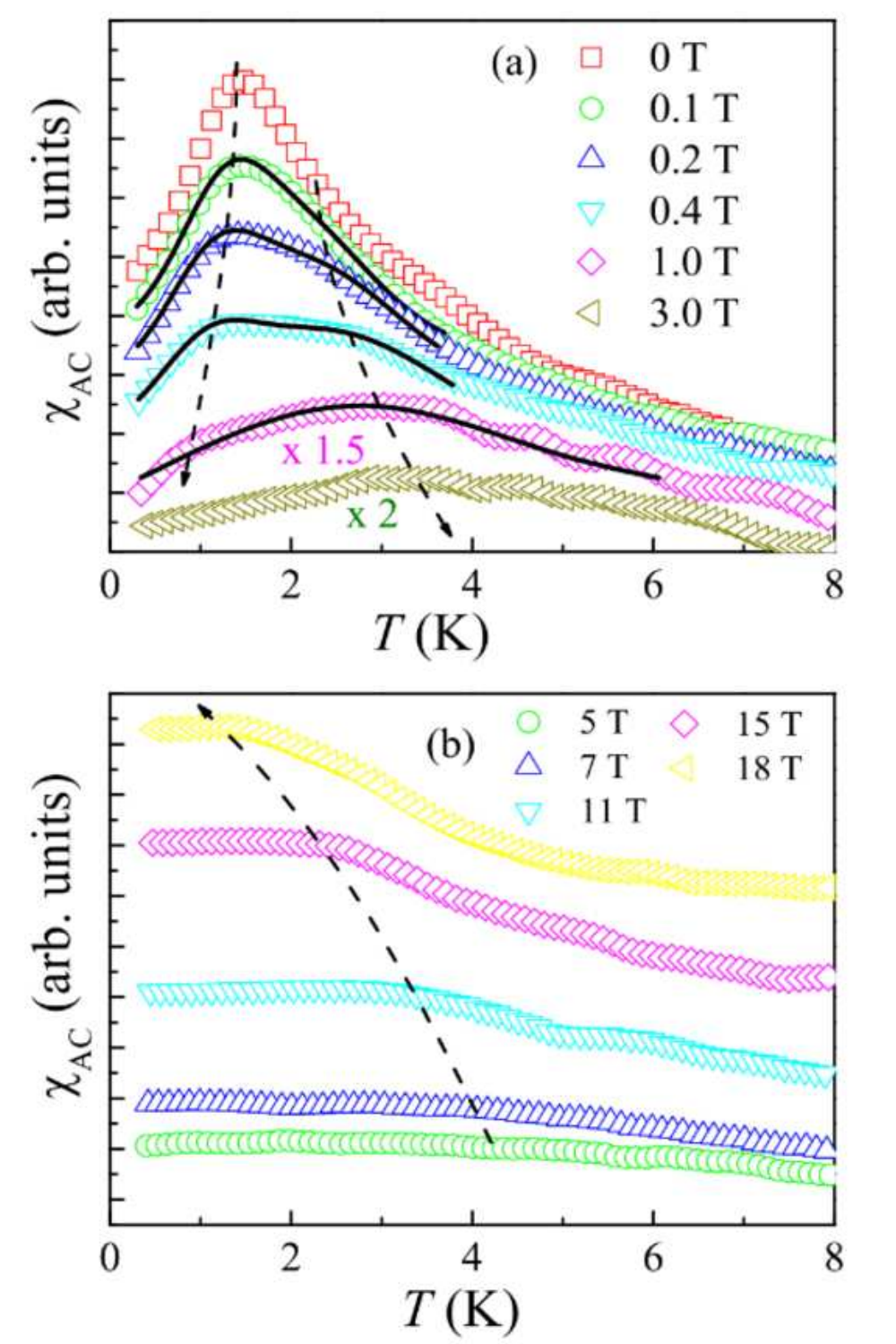}
	\end{center}
	\par
	\caption{\label{Fig:4} (color online) The AC susceptibility for Ba$_8$MnNb$_6$O$_{24}$ under (a) low DC magnetic fields from zero to 3.0 T. Fitting of the susceptibility with two Gaussian peaks is shown for several data sets (solid lines). (b) high DC magnetic fields from 5~T to 18~T. The dash arrowed lines are guide to the eye to highlight the evolution of the peaks' position with increasing DC fields.
}
\end{figure}
\begin{figure}
	\linespread{1}
	\par
	\begin{center}
		\includegraphics[width= 3.2 in]{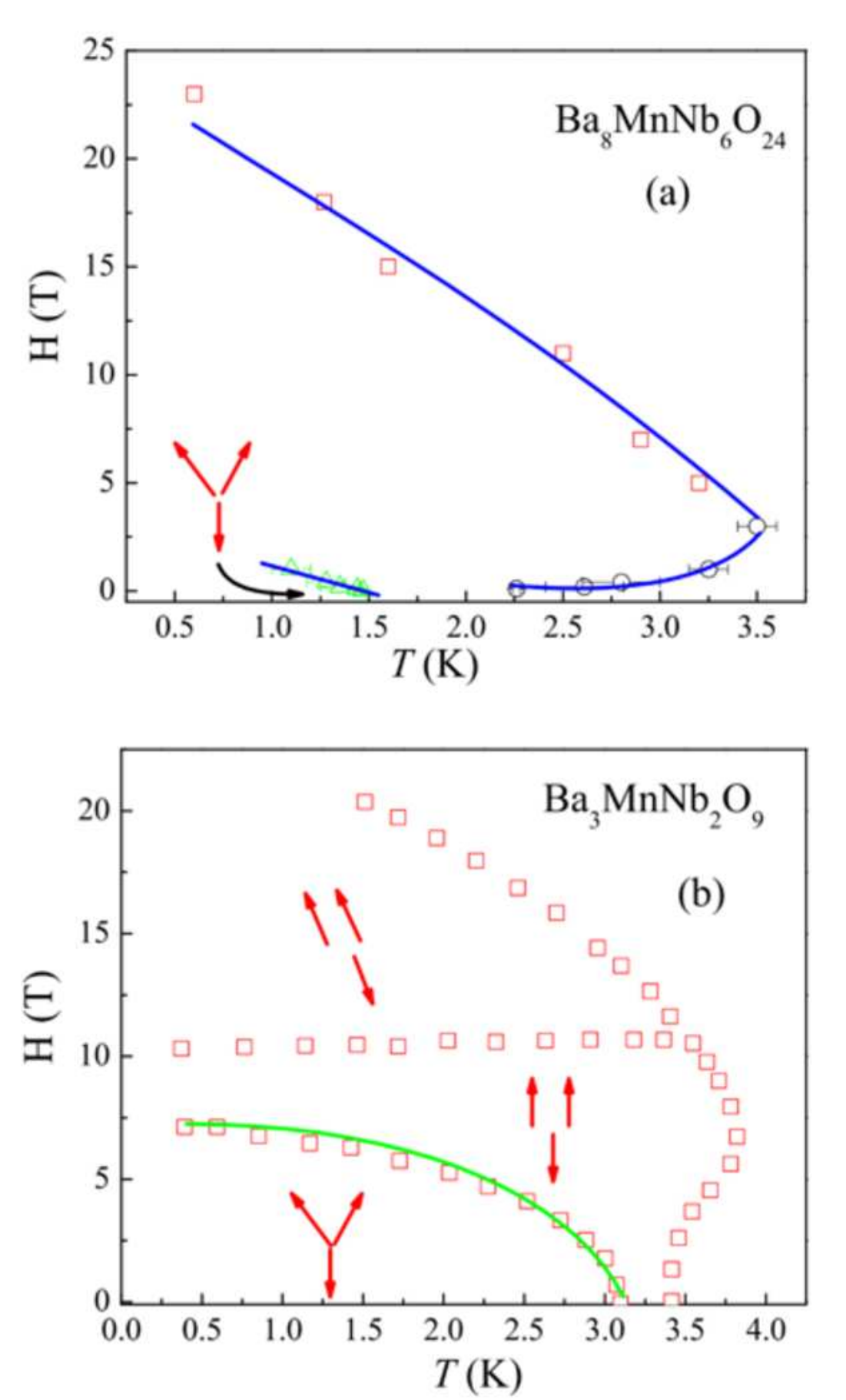}
	\end{center}
	\par
	\caption{\label{Fig:5} (color online) The magnetic phase diagram (a) obtained for Ba$_8$MnNb$_6$O$_{24}$ in this work and (b) obtained for Ba$_3$MnNb$_2$O$_{9}$ in Ref.~\citenum{MnNb}. The drawn spin structures represent the canted 120 degree, up-up-down, and oblique phase, respectively. }
\end{figure}

The AC susceptibility, $\chi_{\text{AC}}$, measured down to 0.3 K at a frequency of 347~Hz was used to probe lower temperature magnetic properties of Ba$_8$MnNb$_6$O$_{24}$, as reported in Fig.~\ref{Fig:4}. At zero DC field, $\chi_{\text{AC}}$ shows a sharp peak at 1.45 K, which indicates a transition to a long range magnetic order. Upon application of a small DC magnetic field of 0.1 T, the peak broadens with a possible shoulder around 2.3~K. The susceptibility was fitted by two gaussian peaks to determine the transitions of $T_{N1}$ and $T_{N2}$, Fig. 4a(solid lines). A possible scenario to explain this behavior is that the single magnetic transition at zero DC field evolves into two transitions upon increasing the DC field, with $T_{\text {N1}}$ refering to the low temperature transition and $T_{\text {N2}}$ the high temperature one. With increasing DC field, Fig.~\ref{Fig:4}(a), $T_{\text {N2}}$ shifts to higher temperatures while  $T_{\text {N1}}$ shifts to lower temperatures and ultimately reaches below 0.3K at 3.0~T.  With even larger DC fields (H $>$ 3.0 T), $T_{\text {N2}}$ changes behavior from a maximum above 3.5~K around 4.0T and shifts to lower temperatures with increasing DC field. Using the saturation field obtained from the DC magnetization and the $T_{\text {N}}$'s obtained from the AC susceptibility data, we draw a magnetic phase diagram, Fig~\ref{Fig:5}(a), for Ba$_8$MnNb$_6$O$_{24}$ and compare it to recent results for Ba$_3$MnNb$_2$O$_{9}$~\cite{MnNb}. A  more detailed discussion of this phase diagram will be presented below.

\subsection{Specific heat}
\begin{figure}
	\linespread{1}
	\par
	\begin{center}
		\includegraphics[width= 3.2 in]{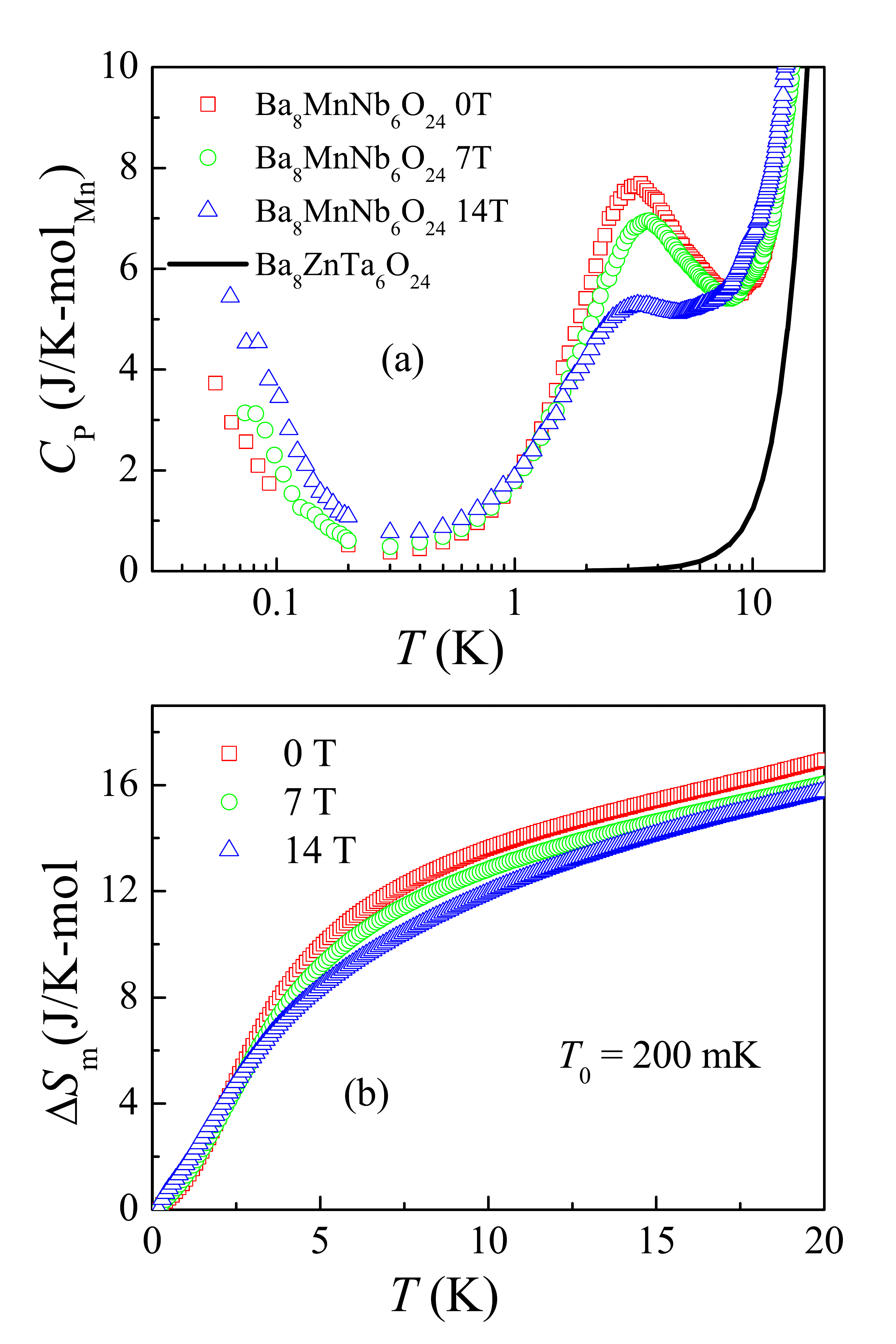}
	\end{center}
	\par
	\caption{\label{Fig:6} (color shown online) (a) The specific heat, $C_{\text{p}}$, of Ba$_8$MnNb$_6$O$_{24}$ at 0 T, 7.0 T, and 14.0 T, and non-magnetic analogue Ba$_8$ZnTa$_6$O$_{24}$ (black line) to expose the lattice contribution. (b) Magnetic entropy change calculated from the integral of $C_{\text{p}}/T$ with lattice subtracted starting at 200 mK.
}
\end{figure}

In Fig.~\ref{Fig:6}(a) we show the specific heat, $C_{\text{p}}$, of Ba$_8$MnNb$_6$O$_{24}$ measured in magnetic fields of 0 T, 7 T, and 14 T. At zero field, $C_{\text{p}}$, shows a broad peak around 4 K but no significant feature matching the kink observed around 1.45 K in the temperature dependence of the AC susceptibility. Moreover, the data shows a sharp increase below 200 mK, which we ascribed to the nuclear Schottky anomaly of the Mn ions, since the naturally abundant 55Mn has nuclear spin $5/2$. This can also explain why the upturn shifts to higher temperatures with higher magnetic fields. Because the magnetic ordered moment increases with the magnetic field and hence opens up the nuclear spin levels through the hyperfine coupling, which leads to the transition at higher temperatures. With increasing fields, the broad peak is somewhat suppressed. We isolate the magnetic contribution to the specific heat, ($C_{\text{m}}$), by subtracting the specific heat of the isostructural non-magnetic compound Ba$_8$ZnTa$_6$O$_{24}$. This yields the magnetic entropy change, $\Delta S_{\text{m}}$, calculated by integrating $C_{\text{m}}$/$T$. For simplicity, we did not subtract the nuclear Schottky anomaly from the data but performed the integration from 200 mK since at this temperature, the $C_{\text{p}}$ is already pretty small. The obtained $\Delta S_{\text{m}}$ is around 16 J/mol-K, a value near to the ideal value for $S=5/2$, which is Rln(6) = 14.9 J/mol-K with R as gas constant.

\subsection{Magnetic structure and excitations}

\begin{figure}
	\linespread{1}
	\par
	\begin{center}
		\includegraphics[width= 3.6 in]{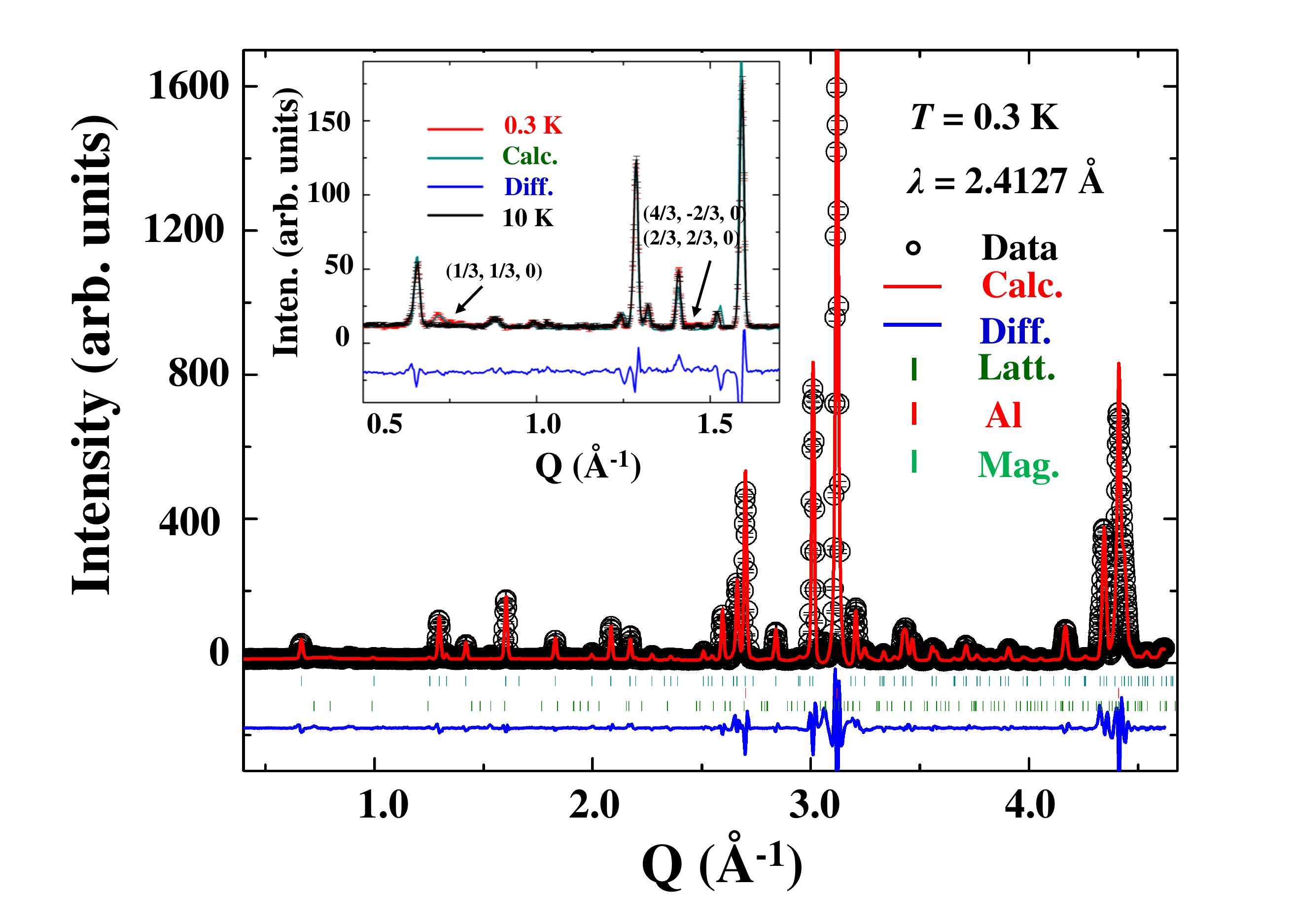}
	\end{center}
	\par
	\caption{\label{Fig:7} (color shown online)  Rietveld refinement of the neutron powder diffraction pattern measured at $T$ = 0.3 K with $\lambda$ = 2.4127  \AA\ for Ba$_8$MnNb$_6$O$_{24}$. Black circles are the measured intensity, the red line is the calculated intensity, and the blue line is the difference. Tick marks are lattice Bragg peak, aluminum and  magnetic Bragg peak positions, from top to bottom, respectively. Insert: Highlight of the difference between the 0.3 K and 10 K neutron powder diffraction patterns to show the magnetic Bragg peak positions.
}
\end{figure}

In Fig.~\ref{Fig:7} we show the NPD pattern of Ba$_8$MnNb$_6$O$_{24}$ measured at 0.3 K with wavelength $\lambda$ = 2.4127 \AA. The data reveals the presence magnetic Bragg peaks that can be indexed by ${\bf Q} = {\boldsymbol{\tau}}+{\bf k}_m$ where ${\boldsymbol{\tau}}$ is a reciprocal lattice vector and ${\bf k}_m=(1/3,1/3,0)$. In the insert of Fig.~\ref{Fig:7}, we highlight the difference between the 0.3 K and 10.0 K data, what reveals the details of these magnetic Bragg peaks. The refined magnetic structure for the above propagation vector ${\bf k}_m$ is shown in Fig.~\ref{Fig:1}(c): it corresponds to a 120$^{\circ}$ spin-structure in the $ab$-plane with spins arranged in a collinear ferromagnetic arrangement between nearest neighbor layers. The refined ordered moment is 4.6 (1) $\mu_B$ for each Mn$^{2+}$ ion. Due to the small number of sizeable magnetic Bragg peaks, the refinement cannot tell whether the spins lie in the $ab$ plane or in a plane containing the $c$ axis (an easy-axis type 120$^{\circ}$ structure).

\begin{figure*}
	\begin{center}
		\includegraphics[width= 6.8 in]{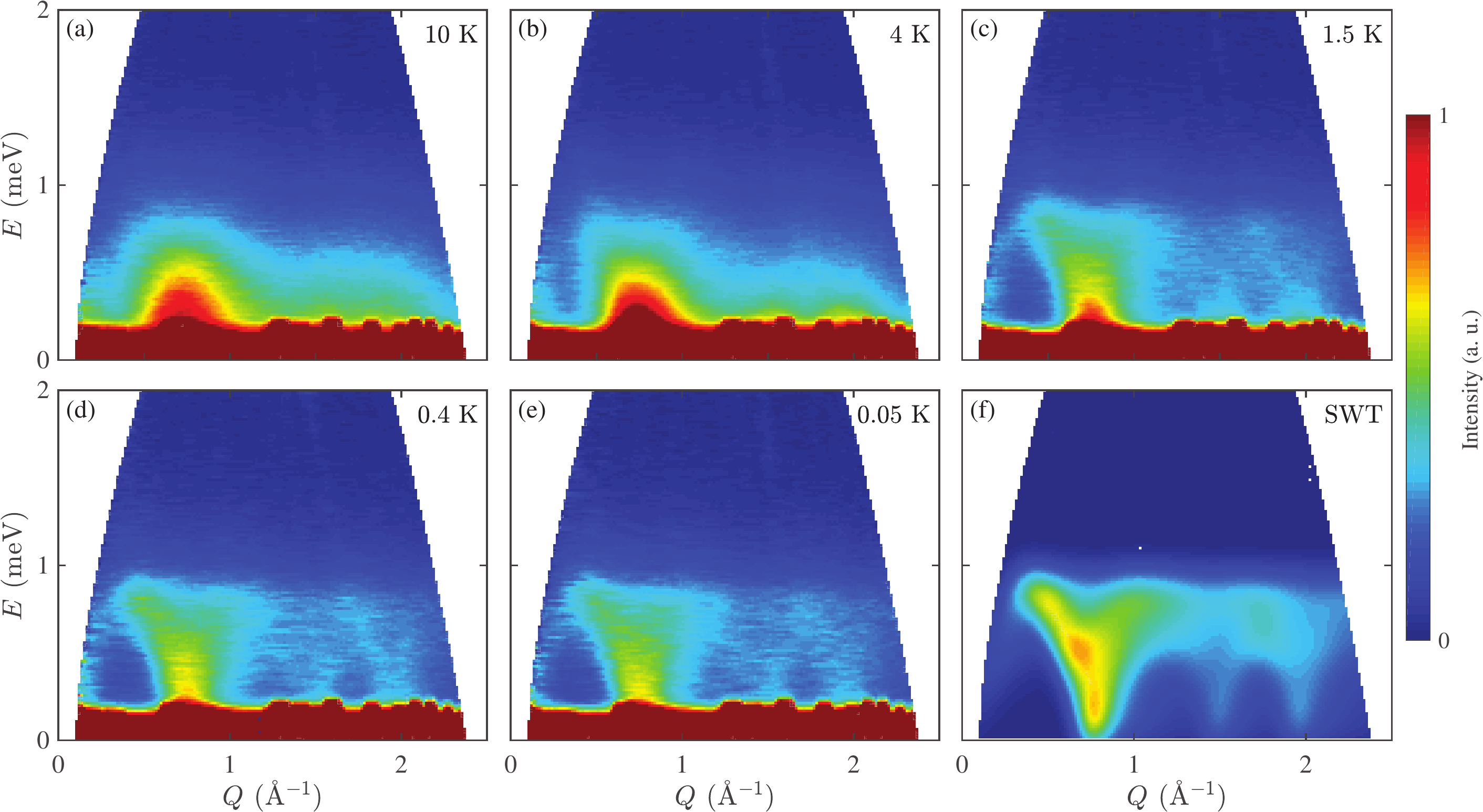}
	\end{center}
	\caption{\label{Fig:8} (color online)  (a-e) Powder inelastic neutron scattering spectra for Ba$_8$MnNb$_6$O$_{24}$ measured at nominal temperatures of 10 K, 4 K, 1.5 K, 0.4 K, and 0.05 K, respectively. (f) The neutron scattering intensity calculated for $J$ = 0.11 meV using linear spin wave theory.
}
\end{figure*}

In Fig.~\ref{Fig:8} shows the powder INS spectra of Ba$_8$MnNb$_6$O$_{24}$ measured at  10 K, 4.0 K, 1.5 K, 0.4 K and 0.05 K, respectively. The INS intensity as a function of momentum transfer $Q$ and energy transfer $E$ allows to track the development of magnetic correlations upon lowering $T$. At high temperatures such as 10 K and 4 K, the INS intensity shows a clear short-range magnetic signal with a momentum dependence peaked  at $Q\!\approx\!0.7$ \AA$^{-1}$.  Upon decreasing temperature and approaching $T_{\text{N}}$ (1.45 K), sharper ridges of intensity emerge from $Q\!\approx\!0.7$ \AA$^{-1}$ with less intense repetitions at $1.5$ \AA$^{-1}$ and $2.0$ \AA$^{-1}$, in perfect agreement with the powder diffraction results. The energy dependence of the main signal reveals gapless excitations extending up to 1.0 meV. These features change very little below $T_\mathbf{N}$ from 0.4 K to 0.05 K. This further confirms the occurrence of a well correlated magnetic state below 1.0K, consistent with long-range magnetic order.

\begin{figure}
	\linespread{1}
	\par
	\begin{center}
		\includegraphics[width= 3.2 in]{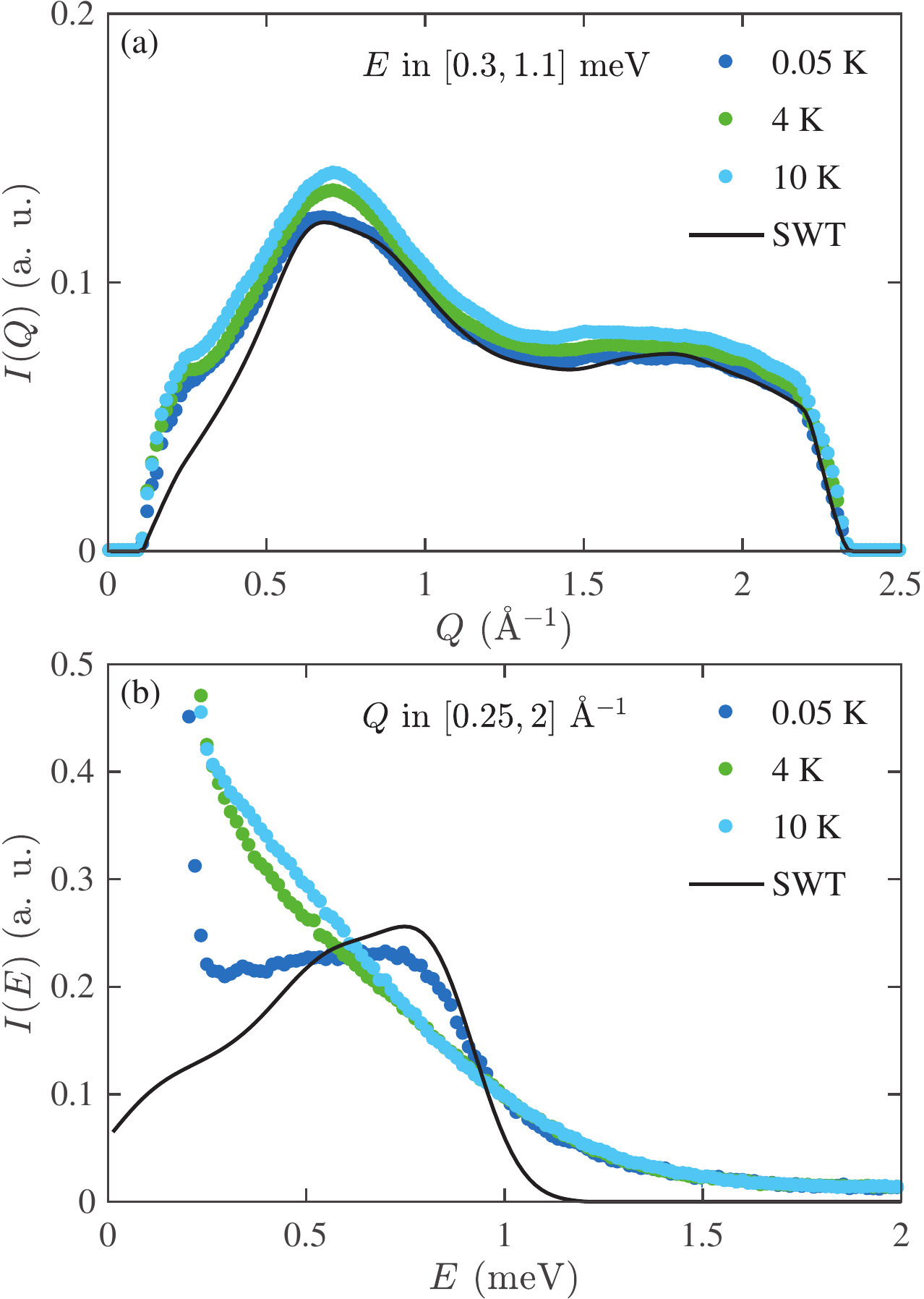}
	\end{center}
	\par
	\caption{\label{Fig:9} (color online) Comparisons between experiment (dots)and linear spin wave theory (black line) as energy-integrated (0.3 $\leq$ E $\leq$ 1.1  meV) and momentum-integrated (0.25 $\leq$ Q $\leq$ 2 \AA$^{-1}$) cuts, respectively.
}
\end{figure}

To model the dynamic magnetic correlations in the ordered state, we resort to linear spin-wave theory at zero temperature~\cite{spinw}. We use a canonical Heisenberg Hamiltonian (Eq.~(\ref{eq:ham})) with the 120$^\circ$ magnetic structure as ground-state. Due to the powder averaging effect, information about possible exchange anisotropies, which we expect to be relatively small, cannot be accurately extracted from the INS data and therefore are not considered here. The best match between the experimental data measured at $T = 0.05$~K and the simulation,  Fig.~\ref{Fig:8}(f), is achieved with the nearest-neighbor exchange interaction $J\!=\!0.11$~meV (or $1.28$~K). This value is consistent with the $J = 1.22$~K calculated from the Curie-Weiss temperature. As shown in Fig.~\ref{Fig:8}(e) and Fig.~\ref{Fig:8}(f), the calculated spectrum reproduces the main features of the experimental data, such as the positions of the zone centers and bandwidth of the magnetic excitations. A more detailed illustration of the good match between experimental and our model is evident in $E$-integrated [Fig.~\ref{Fig:9}(a)] and $Q$-integrated [Fig.~\ref{Fig:9}(b)] cuts. We may also estimate the upper bound of the easy-plane anisotropy $\Delta$ from the INS data. Although such anisotropy does not gap the entire spin-wave dispersion relation in the Brillouin zone, there will be major intensity shifted up in the calculated spin dynamical structure factor at the ordering wave vector. This shift is essentially the energy gap of the out-of-plane mode. Within the linear spin-wave theory, the gap $\Delta\varepsilon$ is proportional to:
\begin{align}
    \Delta\varepsilon = 3JS\sqrt{3(1-\Delta)/2}.
\end{align}
Meanwhile, the inelastic neutron scattering data we have has incoherent elastic line that extends to roughly $0.2$ meV. So we cannot resolve any potential gap if it is smaller than that. Now we just plug the values $J = 0.11$ meV, $S = 5/2$ into the above equation. We can obtain that $\Delta > 0.96$.

\section{DISCUSSION}

A noteworthy feature of Ba$_8$MnNb$_6$O$_{24}$ is that the heat-capacity $C_{\text{p}}$ shows no clear sign of long range magnetic ordering but a broad peak around 4 K.  Previous quantum Monte Carlo studies on quasi-2D antiferromagnetic Heisenberg models have shown that the onset of long-range magnetic ordering is accompanied with a sharp peak in $C_{\text{p}}$ even for inter-layer exchange interactions as small as $J^{\prime}/J$ = 2$\times$10$^{-4}$, see Ref.~\onlinecite{2g}. Upon further decreasing the inter-layer coupling, the sharp peak disappears and only a broad peak remains. Therefore, we believe the sole broad peak in $C_{\text{p}}$ hints at the almost ideal two-dimensional nature of magnetism in Ba$_8$MnNb$_6$O$_{24}$. At the same time, the broad peak indicates that the short-ranged spin correlations have already developed at temperatures higher than $T_\mathrm{N}$. This is consistent with the INS observation, which shows that broad magnetic signals have already developed at as high as 10 K.

In spite of its almost ideal two-dimensional magnetism, Ba$_8$MnNb$_6$O$_{24}$ still appears to order at $T_{\text{N}}$ = 1.45 K with a 120 degree ordering structure as confirmed by the AC susceptibility and neutron diffraction experiments. Since $T_{\rm N}$ increases logarithmically in the inter-layer interaction or in the exchange anisotropy \cite{TN1,TN2,TN3,TN4,TN5,Kawamura1984, Kawamura1985}, we conjecture that the magnetic transition of Ba$_8$MnNb$_6$O$_{24}$ is most likely driven by an easy-plane anisotropy.

It is instructive to compare the magnetic properties of Ba$_8$MnNb$_6$O$_{24}$ with those of the related quasi-2D compound Ba$_3$MnNb$_2$O$_{9}$. For this purpose, the magnetic phase diagram for Ba$_3$MnNb$_2$O$_{9}$ is reproduced in Fig.\ref{Fig:5}(b). The major differences between these two phase diagrams are as follows. First, a two-step transition at $T_{\text{N1}}$ = 3.4 K and $T_{\text{N2}}$ = 3.0 K occurs in Ba$_3$MnNb$_2$O$_{9}$, which indicates that its easy-axis anisotropy. This is not only the normal behavior for Mn$^{2+}$ ions on octahedral sites as for Rb$_4$Mn(MoO$_4$)$_3$, \cite{RbMn} but also for the distorted triangular lattice in $A_3$NiNb$_2$O$_9$ \cite{LuPRB2018} and NaCrO$_2$ (NiGa$_2$S$_4$) \cite{Kawamura2010}. For some special case, the TLAF could exhibit easy-plane anisotropy. For example, the recently studied TLAF Ba$_2$La$_2$MnW$_2$O$_{12}$ shows a single step transition, in which the competition between the antiferromagnetic Mn-O-O-Mn and ferromagnetic Mn-O-W-O-Mn could be the reason for the easy-plane anisotropy. In contrast, Ba$_8$MnNb$_6$O$_{24}$, potentially with a weak easy-plane anisotropy with transferring from one magnetic transition at 0 T to two transitions at very low field (0.1 T), sits closer to the limit of Heisenberg spins. Second, the magnetic phase diagram for Ba$_3$MnNb$_2$O$_{9}$ evolves form the zero field 120 degree ordering, to canted 120 degree ordering, UUD phase, and then oblique phase with increasing applied field, while there is no indication for the existence of the UUD phase in Ba$_8$MnNb$_6$O$_{24}$. Whether this disappearance of UUD phase is intrinsic related to dimensional reduction in Ba$_8$MnNb$_6$O$_{24}$ or is extrinsic due to the polycrystalline sample nature (in which the powder average effect smears the phase boundaries) needs to be clarified by more studies on putative single crystalline samples in the future.

Finally, we compare Ba$_8$MnNb$_6$O$_{24}$ with Ba$_8$CoNb$_6$O$_{24}$, where the main difference lies in the spin quantum number. With $S = 1/2$ magnetic moments, Ba$_8$CoNb$_6$O$_{24}$ is subject to stronger quantum fluctuations and thus exhibits no long-range magnetic order down to $T = 60$ mK and calls for $1/S$ spin-wave theory to model its magnetic excitations. Meanwhile, Ba$_8$MnNb$_6$O$_{24}$, with $S = 5/2$, is essentially in the classical limit and the resulting spin dynamics are well described by linear spin-wave theory without considering any form of magnon-magnon interactions. On top of quantum effects, thermal fluctuations should play an influential role in both systems. Therefore, Ba$_8$MnNb$_6$O$_{24}$, free from strong quantum fluctuations, may serve as a good candidate to investigate the role of thermal fluctuations.

\section{CONCLUSION}

We presented a detailed experimental study of the triangular lattice anitferromagnet Ba$_8$MnNb$_6$O$_{24}$ with $S=5/2$ Mn$^{2+}$ ions forming equilateral triangular lattices. Our results reveal that despite the almost ideally 2D nature of the magnetism and the likely vanishing inter-layer interaction, long-range magnetic order develops at $T_{\text{N}}$ = 1.45 K. Specific heat measurements along with an inelastic neutron scattering study show that short ranged spin correlations are well formed at at temperature as high as $T = 10$ K. Linear spin-wave theory simulations yield a nearest neighbor interaction around 0.11 meV, in good agreement with an estimate from magnetic susceptibility measurements. By comparing with the related triple-perovskite Ba$_3$MnNb$_2$O$_{9}$ and the isostructural Ba$_8$CoNb$_6$O$_{24}$ with effective spin-1/2 Co$^{2+}$ ions, we elucidated the subtle role played by quantum spin number and putative weak anisotropies to produce long-range magnetic ordering in the 2D triangular lattice Heisenberg antiferromagnet Ba$_8$MnNb$_6$O$_{24}$.

\begin{acknowledgments}
J.M. thanks the support from the Chinese Spallation Neutron Source (CSNS) User Special Grant, the NSF China(11774223) and the Ministry of Science and Technology of China (2016YFA0300500). R. R., Q. H. and H.D.Z. thank the support from NSF-DMR through Award DMR-1350002. The work at Georgia Tech (L. G. and M. M.) was supported by the National Science Foundation through Grant No. NSF- DMR-1750186. This research used resources at the High Flux Isotope Reactor, a DOE Office of Science User Facility operated by the Oak Ridge National Laboratory. A portion of this work was performed at the NHMFL, which is supported by National Science Foundation Cooperative Agreement No. DMR-1157490 and the State of Florida. E. S. Choi and M. Lee acknowledge the support from NSF-DMR-1309146.
\end{acknowledgments}

\bibliographystyle{apsrev}

\end{document}